\documentclass[a4paper,10pt,twoside]{cpc-hepnp}
\usepackage{upgreek,fancyhdr}
\usepackage{multicol}
\usepackage{multirow}
\usepackage{graphicx}
\usepackage{booktabs}
\usepackage{amssymb,bm,mathrsfs,bbm,amscd}
\usepackage[tbtags]{amsmath}
\usepackage{lastpage}

\begin{document}

\fancyhead[c]{\small Submitted to Chinese Physics C}
\fancyfoot[C]{\small 010201-\thepage}

\footnotetext[0]{Received 24 Nov. 2016}

\title{A study of raining influence on the environmental radiation background spectra
 with HXMT/HE\thanks{Supported by the National Program on Key Research and Development Project(Grant No.2016YFA0400803)}}

\author{%
      Xu-Fang Li$^{1;1)}$,\email{lixufang@ihep.ac.cn}%
\quad Cong-Zhan Liu$^{1;2))}$,\email{liucz@ihep.ac.cn}
\quad Yi-Fei Zhang$^{1}$,
\quad Zheng-Wei Li$^{1}$,
\\ Xue-Feng Lu$^{1}$, 
\quad Jian-Ling Zhao$^{1}$,
\quad Chang-Lin Zou$^{1}$,
\quad Yu-Peng Xu$^{1}$,
\\ Fang-Jun Lu$^{1}$
}

\maketitle

\address{%
$^1$ Key Laboratory of Particle Astrophysics, Institute of High Energy Physics , CAS, Beijing 100049, China\\
}

\begin{abstract}
Full functional and performance tests were performed many times before the Hard X-ray Modulation Telescope (HXMT) launch. During one of the tests, the count rate curves of the 18 High Energy Detectors £¨HED£©have been found increased consistently within an interval of time. A further study on the correlation between the count rate and rainfall was carried out£¬and the increased net spectrum was also analyzed. The analysis results indicate that the short-lived $^{222}$Rn  decay products £¨$^{214}$Pb and $^{214}$Bi£©in rainwater were responsible for  the transient changes of the background radiation spectra in HEDs. The results show that the HXMT/HEDs have a  good detection sensitivity on X/$\gamma$-rays, and the detector calibration results are effective.
\end{abstract}



\footnotetext[0]{\hspace*{-3mm}\raisebox{0.3ex}{$\scriptstyle\copyright$}2013
Chinese Physical Society and the Institute of High Energy Physics
of the Chinese Academy of Sciences and the Institute
of Modern Physics of the Chinese Academy of Sciences and IOP Publishing Ltd}%

\begin{multicols}{2}

\section{Introduction}

The Hard X-ray Modulation Telescope is the first fully independently developed astronomical satellite of China devoted to X-ray observations in a broad band(1-250keV) ~\citep{lab1}. In December 2015, delivery and integration of the payload flight models were completed. In the consequent nine months, full functional and performance tests were performed many times on the satellite and payloads. In one test in August 2016, the count rates of all the 18 High Energy Detectors (HED) increased obviously within several hours and then dropped to normal level. Many studies found that the concentration of $^{222}$Rn progeny $^{214}$Pb and $^{214}$Bi in precipitation could cause a noticeable increase in environmental gamma-ray intensity \cite{lab2,lab3,lab4,lab5}. The nuclides $^{214}$Pb and $^{214}$Bi decay with short half-lives of 27min and 20min, respectively, and gamma rays of 352keV, 295keV, and 242keV are emitted from lead, and 1764keV,1120keV and 609 keV from bismuth. In this study, rainfall rates recorded by two closed weather stations are compared with HED count rates, and the two spectra during and after the rain are also analyzed. The results verify that the rain is responsible for the count rate increase in HEDs. This work shows that the HXMT/HE has high sensitivity to X/$\gamma$ ray observations, and the results of the detectors ground-based calibration are credible.

\section{Experiment description}

\subsection{High energy detector}

The high energy X-ray telescope (HE) of HXMT consists of a detector array of eighteen independent NaI(Tl)/CsI(Na) phoswich scintillators \cite{lab1,lab6}, and the total geometric area is about $5100cm^2$. The crystal NaI(Tl) is in the front and the thickness is $3.5mm$, while the thickness of CsI(Na) is $40mm$. The phoswich is coupled to a 5 inch photomultiplier tube (PMT) through a $10mm$ thick quartz. At normal operating mode, the NaI(Tl) acts as the X-ray detector, while the CsI(Na) scintillator acts as an active shield to suppress background. Because of the different fluorescence decay time of the two crystals, the signal pulse widths generated by the X-ray photons in the two crystals are different. With the technique of pulse shape discrimination (PSD),it is easy to decide which crystal the X-ray event deposits its energy in. The energy responses of the two crystals of each HED were calibrated using X-ray facility and various radioactive sources before delivery.

\subsection{Testing process}

From 16th to 19th August 2016, a test simulating the satellite operating mode in the orbit was carried out and lasted for 100 hours. The payloads were powered up at noon on 17th August, subsequently, the satellite was simulated to enter the South Atlantic Anomaly (SAA) nine times and the PMT high voltage of each detector was set to $0V$ automatically during the SAA area. All of the payloads were powered off at midnight on August 18th until 19th morning. Except for the time in the SAA, the operating parameters of each HED were kept unchanged during the test.

The count rates recorded by HED-1 in this work as a function of time are shown in Fig.~\ref{fig1}. The background count rate is generally expected to be constant, but there were obvious deviations from the average on 18th August, and which came back to normal a few hours later. The background data was reduced to zero ten times corresponding to the events entering the SAA nine times and payloads shutdown. The count rate responses of the other 17 HEDs are consistent with that of HED-1.

\end{multicols}
\begin{center}
\includegraphics[width=12cm]{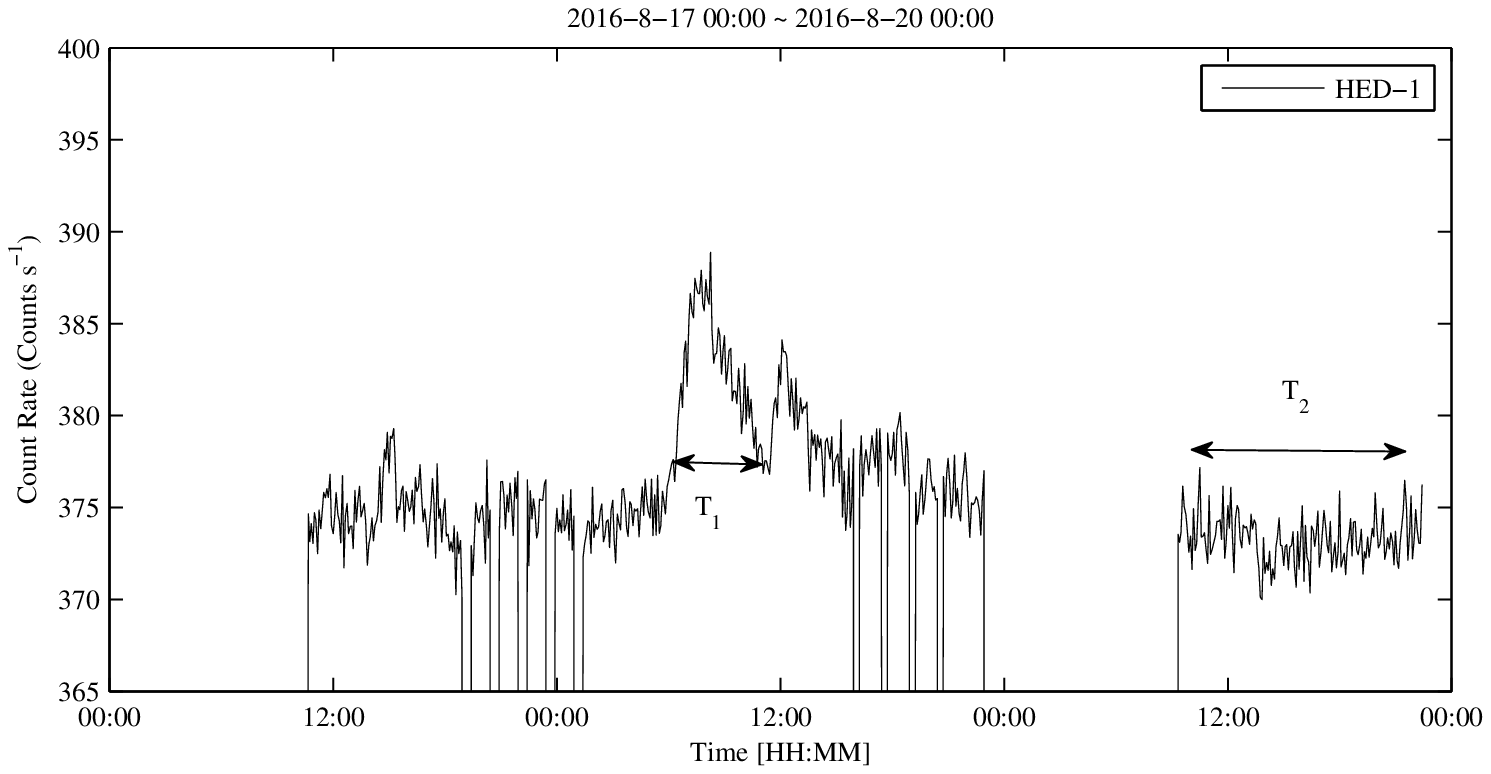}
\figcaption{\label{fig1} Count rate curve of $HED-1$ background response. Since the payloads powered up at noon on Aug. 17th, the background data was reduced to zero 10 times corresponding to the events entering the SAA nine times and payloads shutdown. The count rate increased significantly in the morning of Aug. 18th and returned to normal at that evening.}
\end{center}

\begin{multicols}{2}

\section{Data analysis and discussion}

\subsection{Correlation between rainfall and HED count rate}

 The rainfall data from two meteorological stations closest to the test site was used for the analysis. The two stations are Beijing WenQuan Station (W.Q.) which is $8km$ southwest away from the test site, and Beijing ShangZhuang station (S.Z.) which is 5.3km northwest away from the test site. The data from  the 18 HEDs were re-binned with a $5min$ step, and the rainfall rates from the two station were also re-binned with a $5min$ step. Fig.~\ref{fig2} shows the correlation of HE counts and the rainfall data, and the gamma background increased significantly during rain occurrence. Because the stations are a few kilometers west of the laboratory, and rain clouds moved from northwest to southeast, the evolution of gamma counts detected by HE was slightly delayed than that of rainfall at the two stations.

\begin{center}
\includegraphics[width=8cm]{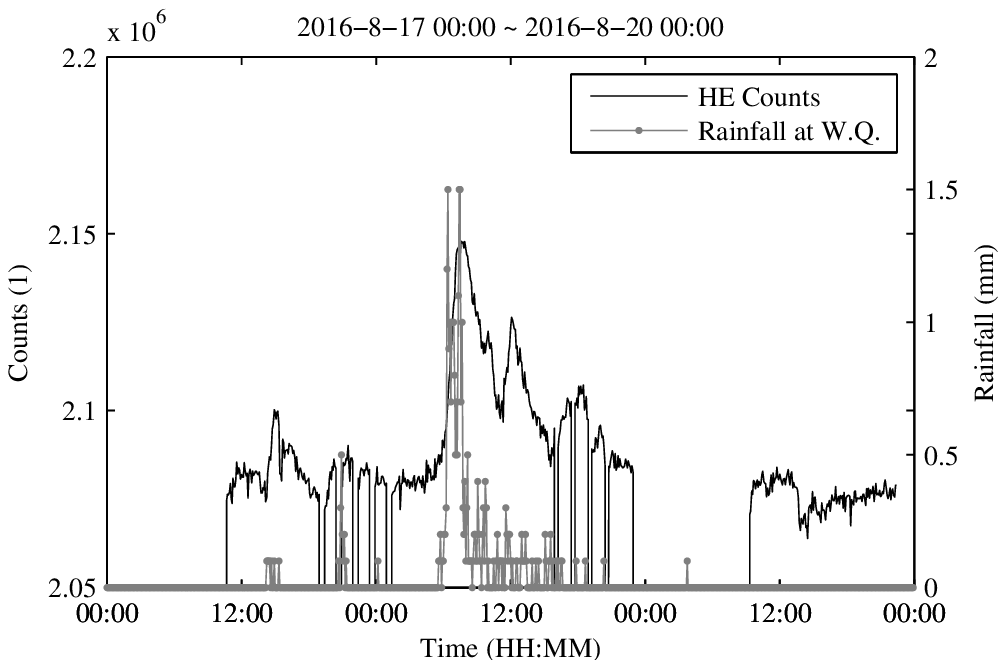}
\includegraphics[width=8cm]{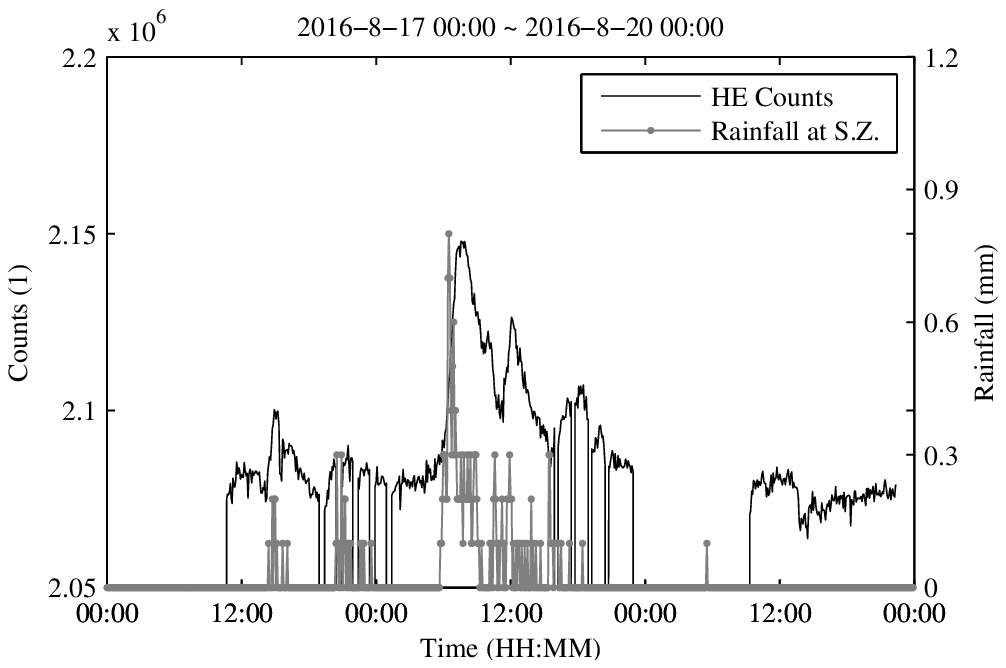}
\figcaption{\label{fig2}   HE counts and rainfall detected at two nearest weather stations as a function of time, it shows a good correlation.}
\end{center}

\subsection{Spectrum analysis}

\subsubsection{Data selection}

To study the background radiation induced by rain, signals collected during the following two periods (shown in Fig.~\ref{fig1}) were chosen and analyzed, time period 1($T_1$) with rain is from $6:40:51$ to $10:29:11$ on August 18th, time period 2($T_2$) without rain is from $10:22:31$ to $22:25:51$ on August 19th. Applying PSD, pulse height spectra recorded by NaI and CsI of the 18 HEDs in the two periods were obtained,respectively. Obviously, the CsI signals dominate the on-ground background, then only the CsI pulse height spectra (mark as $S$) were studied in this work.

\subsubsection{Spectrum alignment}

With the help of multichannel analyzer (MCA), the pulse amplitudes of the incident $\gamma$-ray photons are converted to digital channel number, and the effective number of channels of HED are from 20 to 275. In order to get the best statistical accuracy, the spectra of all HEDs need to be combined into a single spectrum. However, the Energy-Channel relationships of the HEDs are different. For combination of all the HEDs spectra correctly, the raw data must be re-binned and sorted into a new set of channels of an energy spectrum ~\citep{lab7}. The detailed method is described as follows.

1) Sampling was used to convert the discrete pulse height channels to continuous  values that characterizes the source of the pulses. The total number of all pulses in the measured spectrum :
\begin{equation}\label{eq1}
  N_{S}=\sum_{i=20}^{275}N_{S(i)}.
\end{equation}
Where, $N_{S(i)}$ is the number of counts in the $i$th channel of $S$. With uniform sampling on the interval [$i$-0.5£¬$i$+0.5], a set of continuous pulse height values were generated with a number of $N_{S(i)}$. In this way, the pulse height $P_j$ ($j$ is the pulse number from $1$ to $N_{S}$ ) of all radiation events were re-sampled in the continuous interval [$19.5$£¬$275.5$].

2) Using the calibrated channel-energy relationship of that CsI detector $E(P_j)=k\cdot P_j+b$ , the energy of incident photon $E_j$  (keV) was calculated from the pulse height $P_j$ which was obtained from step 1).

3) According to the on-ground calibration results, all the $E_j$  (keV) in the energy range 35 to 965keV were grouped into 93 bins with an energy interval $\Delta E=10keV$ . Then a new spectrum  $S_E$ was generated according to the original spectrum $S$. For the $k$th channel of $S_E$, the energy center of this channel is labeled as $E_{S_{E}(k)}$ , the energy of the channel edges is labeled: $E_{S_{E}(k)_{floor}}$ ($=E_{S_{E}(k)}-\Delta E/2$) and $E_{S_{E}(k)_{ceil}}$ ($=E_{S_{E}(k)}+\Delta E/2$) respectively. Then the recorded count can be calculated from:
\begin{equation}\label{eq2}
  N_{S_E(k))}=\sum_{j=i}^{i+m}p_{j}\cdot N_{S(j)}.
\end{equation}

Where  $$E_{S(i-0.5)}\leq  E_{S_{E}(k)_{floor}}< E_{S(i+0.5)},$$
$$E_{S(i+m-0.5)}\leq  E_{S_{E}(k)_{ceil}}< E_{S(i+m+0.5)},$$
$$p_{j}=\frac{A_{j,k}}{N_{S(j)}},$$
$A_{j,k}$  is number of counts with energy between $E_{S_{E}(k)_{floor}}$ and $E_{S_{E}(k)_{ceil}}$ in the $j$th pulse height interval . $N_{S(j)}$ is the radiation intensity in the $j$th pulse height interval, The process of re-sampling and re-grouping is illustrated in Fig.~\ref{fig3}.

\end{multicols}
\begin{center}
\includegraphics{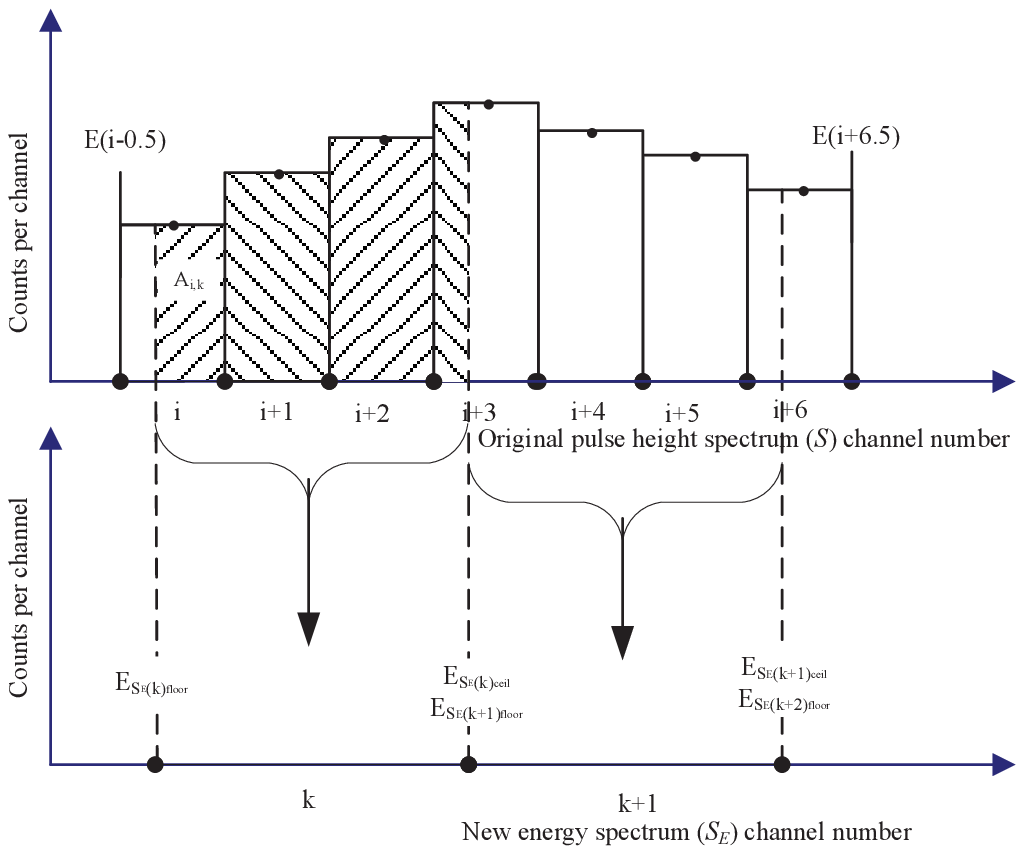}
\figcaption{\label{fig3} Schematic representation of the ¡°re-binning¡± process to relocate pulse height spectrum to energy spectrum. The original pulse height spectrum is shown at the top as a series of points plotted at the midpoints of the channels $i$,$i$+1£¬$i$+2¡­.Histogram was the distribution of uniform sampling to these points. In any given channel interval the area under the distribution is proportional to the number of counts. At the bottom is the new energy channel scale. }
\end{center}

\begin{multicols}{2}

4) Statistical error analysis.
According to the theory of error propagation ~\citep{lab8} and Equation ~\ref{eq2}, the statistics deviation for $N_{S_E(k))}$  can be calculated as:

\begin{equation}\label{eq3}
  \sigma _{N_{S_{E}(k)}}=[\sum_{j=i}^{i+m}p_{j}^{2}\sigma_{N_{S(j)}}^{2} ]^{1/2}.
\end{equation}

Where $\sigma _{N_{S(i)}}=\sqrt{N_{S(i)}}$.

Then the CsI energy spectra of each HED recorded in the time period $T_1$ and $T_2$ have been obtained from the above process. Spectra of 15 HEDs, for which the working voltages during test were the same as that during calibration, were used to form the HE energy spectra $S_{E,total}$ . The total count in the $k$th channel

\begin{equation}\label{eq4}
  N_{S_{E,total}(k)}=\sum _{n=1}^{15}N_{S_{n,E}(k)}.
\end{equation}

The corresponding statistics error
\begin{equation}\label{eq5}
  \sigma _{N_{S_{E,total}(k)}}=[\sum _{n=1}^{15}\sigma _{N_{S_{n,E}(k)}}^{2}]^{1/2}.
\end{equation}
 where $\sigma _{N_{S_{n,E}(k)}}$ is the standard deviation of the count $N_{S_{n,E}(k)}$  of the $n$th HED calculated from Eq.~\ref{eq3}.

We define the total spectra recorded during the time period $T_1$ with rain as $S_{E,1}$ , and define the spectra detected in the period $T_2$ without rain as $S_{E,T_2}$  ,which was time normalization to $T_1$ :$S_{E,2}=\frac{T_1}{T_2}S_{E,T_2}$
 .  Then the difference of the two HE/CsI spectra $S_{E}^{'}=S_{E,1}-S_{E,2}$ is the net spectrum of precipitation, as shown in Fig.~\ref{fig4}. The net count in the $i$th channel of $S_{E}^{'}$  is defined as $N_{S_{E}^{'}(i)}$ , and its standard deviation is :
 \begin{equation}\label{eq6}
  \sigma _{N_{S_{E}^{'}(i)}}=[\sigma _{N_{S_{E,1}(i)}}^{2}+\frac{T_{1}^{2}}{T_{2}^{2}}\sigma_{N_{S_{E,T_2}(i)}}^{2}]^{1/2}.
\end{equation}

\begin{center}
\includegraphics[width=8cm]{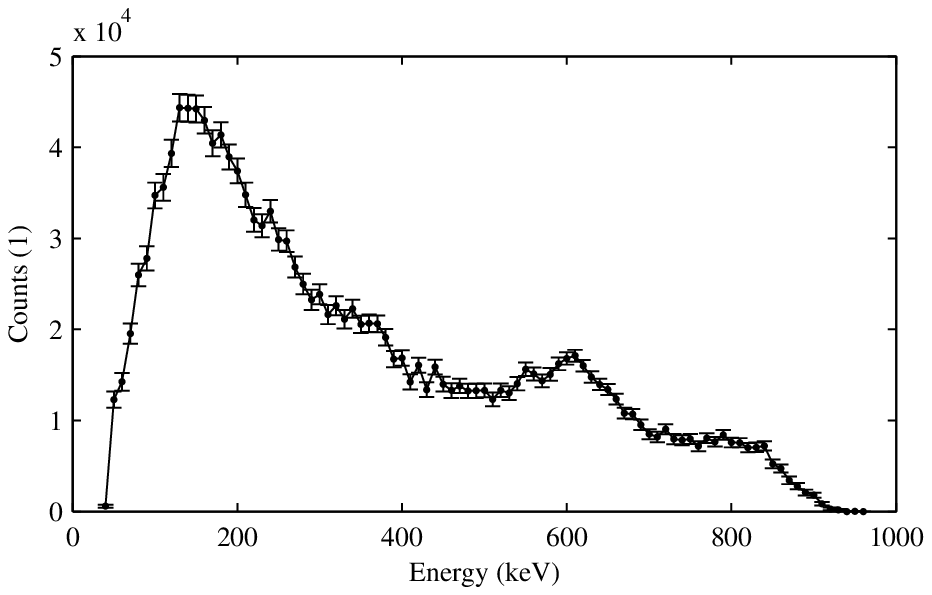}
\figcaption{\label{fig4} Increased environmental gamma-ray spectrum $S_{E}^{'}$  during the rain period $T_1$ measured by HE/CsI detectors.}
\end{center}

\subsubsection{Analysis of spectra with peaks}

The process of spectrum analysis is shown in Fig.5. First of all, smoothing the net spectrum $S_{E}^{'}$  by the moving least squares smoothing formula (Eq.~\ref{eq7}) so as to minimize the effects of statistical fluctuations. The smoothed spectrum is denoted by $S^{''}$ .

\begin{equation}\label{eq7}
  \bar{N_i}=\frac{1}{35}(-3N_{i-2}+12N_{i-1}+17N_{i}+12N_{i+1}-3N_{i+2}).
\end{equation}

 Secondly, the contribution of the the underlying continuum must be subtracted from the smoothed spectrum. Statistics-sensitive Nonlinear Iterative Peak-clipping (SNIP) is known as the optimal baseline subtraction method and has been widely used in X/$\gamma$ spectrum analysis \cite{lab9,lab10,lab11,lab12}. Based on the HE/CsI spectral characteristics, the SNIP algorithm was applied just within the 14th to 80th channel intervals of $S^{''}$ to avoid the count inflection point in the low and high energy regions. The continuum estimated by SNIP is defined as $S_{bg}^{''}$. Channel-by-channel subtraction of the continuum ($S^{''}-S_{bg}^{''}$ ) then produces net data for peak determination.

 Finally, each peak was fitted with Gaussian function. Because the peak1 (as shown in Fig.~\ref{fig5}) is asymmetry, data in energy range 570 to 710keV were chosen to fit. The peak analysis results include centroid, full width at half maximum (FWHM) and the statistical significance of the signals are shown in Table. ~\ref{tab1}. The definition for the statistical significance ~\citep{lab13} is:

\begin{equation}\label{eq8}
  P=[n-(b_{1}+b_{2})]/\sqrt{b_{1}+b_{2}}.
\end{equation}

Where $n$ is the peak gross area in the spectrum $S_{E,1}$ in a prescribed energy limits, $b_1$ and $b_2$ are the summation of counts over the given energy region in the spectrum $S_{E,2}$ and $S_{bg}^{''}$, respectively. So $(b_1+b_2)$ is the total background counts in the prescribed energy region of $S_{E,1}$, and $n-(b_1+b_2)$ is the net peak area in the given region, which also equals the summation of counts in the given channels of the net spectrum ($S^{''}-S_{bg}^{''}$ ).

\end{multicols}
\begin{center}
\includegraphics{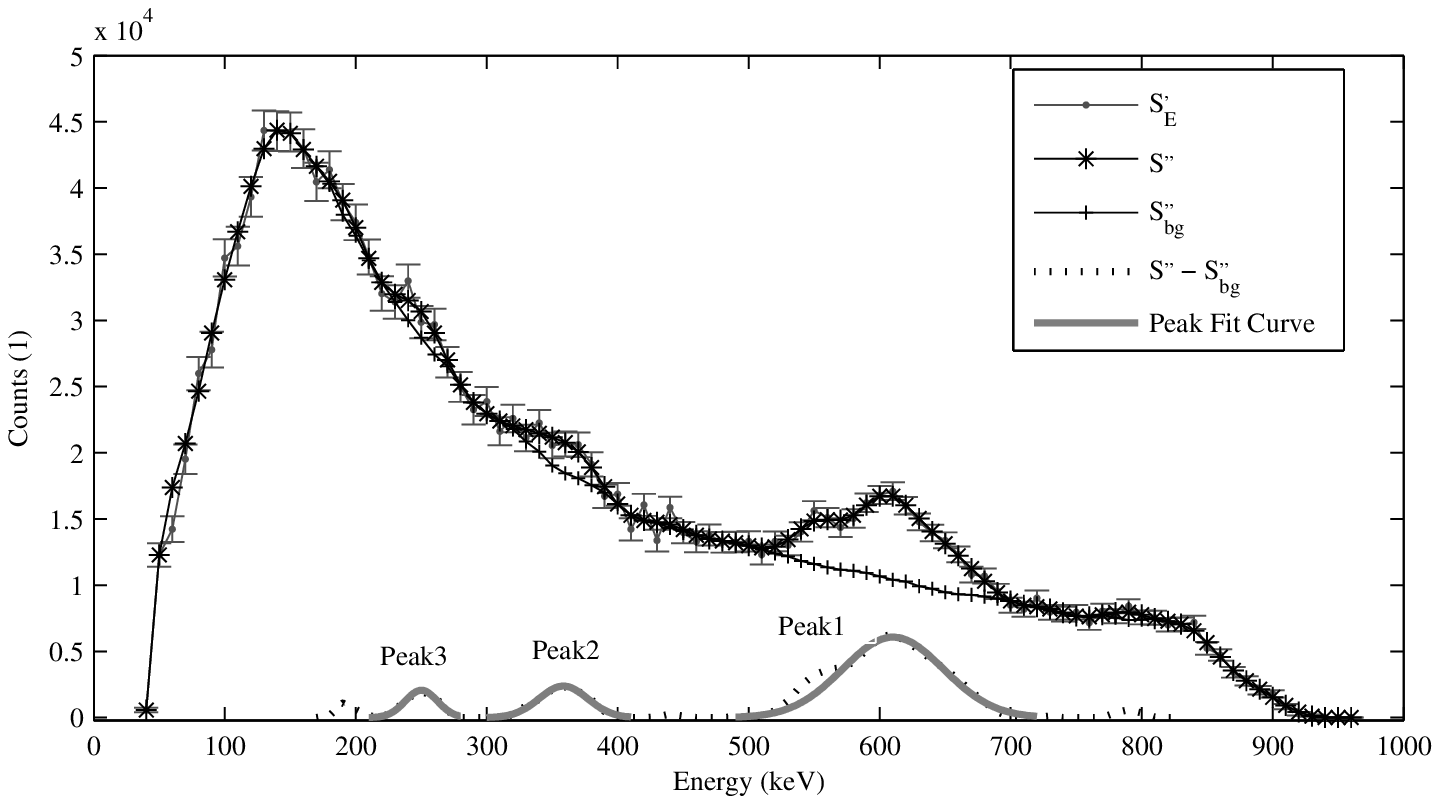}
\figcaption{\label{fig5} The process and results of CsI energy spectrum analysis. $S_{E}^{'}$ is the net spectrum of rainwater, $S^{''}$ is the smoothed spectrum, $S_{bg}^{''}$ is the continuum estimated using the SNIP algorithm, $S^{''}-S_{bg}^{''}$ is the net data for peak determination. All the three peaks were fitted with Gaussian function and the fit curves are plotted in solid.}
\end{center}

\begin{center}
\tabcaption{ \label{tab1}  Peak information and the statistical significance.}
\footnotesize
\begin{tabular}{ccccc}
 \hline
 \multirow{2}{*}{Peak No.} &
 \multirow{2}{*}{Energy} &
 \multirow{2}{*}{FWHM} &
 \multicolumn{2}{c}{Statistical significance of signals} \\
 \cline{4-5}
   & (keV) & (keV)  & Energy interval & statistical significance P \\
 \hline

Peak 1 & $610\pm2.3$ & $90.2\pm6$  &  525-695keV  & 24.2 \\
Peak 2 & $358.5\pm1.5$ & $43.4\pm3.5$  &  315-405keV  &  4.4  \\
Peak 3 & $250\pm1.8$ & $28.9\pm4.2$ &  225-275keV  &  2.7  \\
 \hline
 \end{tabular}
\vspace{0mm}
\end{center}

\begin{multicols}{2}

According to the statistical significance P of the lines, the peak of 610keV and 358.5keV are positively identified as two line emissions from the rainfall precipitate. The peak of 250keV with a significance of 2.7 is suspected to be detected. These three lines are from lead and bismuth: 609keV from $^{214}Bi$ and 352keV, 242keV from $^{214}Pb$. The fitting  error of energy is $0.16\%$£¬$1.85\%$ and $3.31\%$ respectively, less than the fitting error ($<5\%$) of the E-C relationship of CsI.

\section{Conclusions}

Time-correlated data from HEDs and two weather stations proved that the increased counts in background recorded on HEDs are related to the precipitation. Studies of HE/CsI response to rain-induced increases of background radiation indicates that the $\gamma$-ray photons from the $^{214}Bi$ and $^{214}Pb$ were recorded throughout the duration of the rainfall. Consequently the deposition of radon progenies $^{214}Bi$ and $^{214}Pb$  in precipitation contributes to the HXMT/HE increased environmental gamma-ray counts. This work shows that the telescope HXMT/HE is sensitivity to environmental X /$\gamma$ radiation, and the functions and performances of that are reliable. Meanwhile, this work verified that the on-ground calibration results of HXMT/HE are credible.

\vspace{15mm}
\acknowledgments{We appreciate Beijing Meteorological Service for providing the rainfall information.}


\vspace{15mm}


\end{multicols}

\clearpage
\end{document}